\documentclass[conference,compsoc]{IEEEtran}
\ifCLASSOPTIONcompsoc
  \usepackage[nocompress]{cite}
\else
  \usepackage{cite}
\fi
\ifCLASSINFOpdf
\else
\fi
\usepackage{graphicx}
\usepackage{multirow}
\usepackage{pgf-pie}
\usepackage{pgfplots}
\usepackage{listings,xcolor}
\usepackage{pifont}
\usepackage{soul}
\usepackage{amsmath}
\usepackage[plain]{algorithm}
\usepackage{algpascal}
\usepackage[bookmarks=false]{hyperref}
\usepackage{subcaption}
\usepackage{graphicx}

\begin{document}
%
\title{An Automated Vulnerability Detection Framework for Smart Contracts\\
}

\author{\IEEEauthorblockN{Feng Mi, Chen Zhao, Zhuoyi Wang, Sadaf MD Halim, Xiaodi Li\\
Zhouxiang Wu, Latifur Khan, Bhavani Thuraisingham}
\IEEEauthorblockA{Department of Computer Science, The University of Texas at Dallas, Richardson TX, USA \\
\{Feng.Mi, Chen.Zhao, zxw151030, SadafMd.Halim, xiaodi.li, zhouxiang.wu,  lkhan, bxt043000\}@utdallas.edu}
}

%


\maketitle

\begin{abstract}
With the increase of the adoption of blockchain technology in providing decentralized solutions to various problems, smart contracts have become more popular to the point that billions of US Dollars are currently exchanged every day through such technology. 
Meanwhile, various vulnerabilities in smart contracts have been exploited by attackers to steal cryptocurrencies worth millions of dollars. The automatic detection of smart contract vulnerabilities therefore is an essential research problem.
Existing solutions to this problem particularly rely on human experts to define features or different rules to detect vulnerabilities. However, this often causes many vulnerabilities to be ignored, and they are inefficient in detecting new vulnerabilities.  
In this study, to overcome such challenges, we propose a framework to automatically detect vulnerabilities in smart contracts on the blockchain. More specifically, first, we utilize novel feature vector generation techniques from bytecode of smart contract since the source code of smart contracts are rarely available in public.
Next, the collected vectors are fed into our novel metric learning-based deep neural network(DNN) to get the detection result. 
We conduct comprehensive experiments on large-scale benchmarks, and the quantitative results demonstrate the effectiveness and efficiency of our approach.
\end{abstract}


%
\IEEEpeerreviewmaketitle

\section{Introduction}
\label{sec:smart_intro}
Smart contracts are programs or applications that execute on the blockchain in a decentralized way.
With smart contracts, arbitrary computations can be performed in addition to transaction-based systems. Through the blockchain technique, different entities can interact without a centralized authority \cite{ruan2019fine, frey2019dietcoin}. 
Smart contracts can be used all across the chain from financial services to healthcare insurance \cite{griggs2018healthcare,cohn2017smart,buterin2014next}. Due to the rapid growth in their deployments, various vulnerabilities have been exploited by attackers to steal cryptocurrencies worth millions of dollars \cite{hynes2018demonstration}. The requirement of vulnerability detection hence attracts more and more attention in recent times\cite{vscl}. 

Smart contracts are commonly written in a high-level language, such as Ethereum's Solidity, and it is then translated to low-level bytecode for deployment on the blockchain. A blockchain can be seen as a public virtual machine where the bytecode of the smart contracts is easy to get. Using the Application Binary Interface (ABI), users can interact with a deployed smart contract to complete tasks. Various vulnerabilities have been discovered and exploited for financial gain purposes by attackers, which originate from a wide range of issues, including deficient programming methodologies, language design issues, toolchains, and buggy compilers such as those in the Solidity compiler\footnote{https://solidity.readthedocs.io/en/latest/bugs.html}. Some common vulnerabilities are transaction-ordering dependence, timestamp dependence, mishandled exceptions, reentrancy vulnerability, unsecured balance, destroyable contract, and stack-overflow \cite{luu2016making, brent2018vandal}. One challenge to detect the vulnerabilities is
that most smart contracts’ solidity codes are not provided and
people can only get the bytecode of these smart contracts.
This indicates developing tools to detect vulnerabilities using
bytecode is more practical for users.

Currently, there are many blockchains with smart contract capabilities. Ethereum has become the \textit{de facto} standard platform for smart contract development, with a market capitalization of \$66B (USD)\footnote{https://coinmarketcap.com/currencies/ethereum/} and having \$14.3B (USD)\footnote{https://defipulse.com/} cryptocurrencies locked in its smart contracts by the end of 2020. For this reason, we particularly focus on Ethereum smart contracts in this study.


Previous works in smart contract defense focused on discovering vulnerabilities in smart contracts. Oyente\cite{luu2016making}, Mythril\cite{mythril}, Osiris \cite{osir}, TeEther\cite{tether}, and Zeus\cite{kalra2018zeus} performed vulnerability discoveries by leveraging symbolic execution, Z3 solvers \cite{de2008z3} and pre-defined rules. Symbolic execution has a scalability problem because of the time-consuming in execution procedure. On the other hand, rule-based systems rely on human experts to define different rules to detect bugs in the programs. Besides, machine learning approaches \cite{wang2020xg,J2019fuzz} and \cite{gnn} detect vulnerabilities based on source code.   The machine learning frameworks in \cite{wang2020xg,J2019fuzz} fully depend on decisions from other tools. They do not consider such decisions also have false positives and false negatives. \cite{AWDLSTM,escort} utilize sequence model and achieve good performance when their methods apply to certain types of vulnerabilities.

In this paper, we propose a novel framework to address the issues mentioned above. It makes detection on bytecodes which are generally available in public for smart contracts. After converting the bytecodes to a sequence of operation codes (opcodes) by disassembler, we apply the control flow graph (CFG) and depth-first search (DFS) to form a new sequence of operation codes that better reflect the program execution semantics. After sequences encoding, we use a supervised metric learning-based deep neural network (DNN) to analyze these vectors for vulnerability detection. The metric learning approach works as a regularization term on the deep learning model, reducing intra-class variances and improving discrimination in the embedding space. Our framework has its strengths compared with existing approaches: (1) It does not rely on defined rules. It can auto-extract and learn features. (2) Unlike dynamic analysis approaches, it does not need to execute the program multiple times. Our framework only needs to construct feature vectors and feed them into the detection model when smart contracts come. (3) It trains and detects using bytecodes that can be easily and largely acquired. As a consequence, the detection made is more convincing and practical. (4) It does not adopt existing traditional machine learning models. We develop a DNN with metric learning loss for making smart contract vulnerability detection robust. 

In summary, the contributions of the proposed approach are as follows: 

\begin{itemize}

    \item We propose a novel vulnerability detection framework for smart contracts, which constructs CFGs on more readable low-level opcodes to reflect the execution scenario more clearly.

    \item A metric learning-based DNN is developed, which works as an auxiliary term to improve the discrimination on the embedding space. It makes classification more precise by conducting the instance-weight strategy during the model optimization step. Furthermore, this metric learning helps to uncover new types of vulnerabilities on the fly where during training time the classifier may not have any knowledge of these new vulnerabilities.

	\item We empirically evaluate the proposed framework over real-world, large-scale Ethereum smart contract datasets constructed in different ways and compare its performance against other baselines. The results show its effectiveness and efficiency in detecting vulnerability automatically.
    
\end{itemize}

This paper is organized as follows.
Section \ref{sec:smart_related} presents related work.
Section \ref{sec:smart_arc} introduces the stages of the vulnerability detection in our framework.
Section \ref{sec:smart_evaluation} points out the benchmark and discusses results from our empirical analysis experiments.
Section \ref{sec:smart_discussion} discusses limitations of the proposed approach and future works. 
Section \ref{sec:smart_conclusion} draws conclusions.

\section{Related Work}
\label{sec:smart_related}
\subsection{Static Analysis Tools}

Static analysis tools rely on the analysis of smart contracts without executing it. The characteristic of the static tool is it usually performs a high coverage rate. Static methods can cover as many execution paths as possible. Most static analysis tools are rule-based.
One advantage of these methods is the rapid detection speed, which guarantees scalability. However, in the rule-based approaches, if the designed rules do not cover some of the vulnerabilities, it may lead to a high false positive rate and poor accuracy. Vandal \cite{brent2018vandal}, for example, is a static framework that uses low-level bytecode to make smart contracts secure. In this framework, the bytecode from Ethereum smart contracts is converted to higher-level logical relations. Vandal enables users to identify security problems in the contracts by using declarative logic rules to enumerate the problems, resulting in improved security analysis.

\subsection{Dynamic Analysis Tools}

There are a couple of dynamic analysis tools. Some are based on the symbolic execution by constraint solvers, while the others use some test cases as input of the program and analyze the results. 
Mythril and Oyente\cite{luu2016making} are based on the symbolic execution where input for program functions are symbols that represent arbitrary input values. EVM interpreter keeps track of the program states it encounters and collects constraints on inputs from predicates encountered in branch instructions. 
These tools have scalability problems due to long time cost in the execution procedure. 

\subsection{Machine learning-based Tools}

Several works attempt to utilize machine learning methods to perform automated smart contract detection. \cite{wang2020xg,J2019fuzz} perform vulnerability detection using basic machine learning models. However, when decompiling source codes into opcode, this process can cause information loss. \cite{gnn} utilizes graph model to make detection on source codes of smart contracts. Specifically, this work builds a contract graph from the contract's source code where nodes and edges represent critical function calls/variables and temporal execution traces. It can detect three types of vulnerabilities. Our method is a more generalized framework that can detect various vulnerabilities. For example, in the CodeSmell dataset containing 20 types of vulnerabilities, our method achieves good results.  
Furthermore, all the source codes frameworks \cite{wang2020xg,J2019fuzz,gnn} uses much smaller datasets than ours due to the reason that source codes are not generally available compared with the bytecode. Generally, the bytecode of the smart contracts is available as long as the smart contract is deployed in the blockchain.   \cite{escort}, and \cite{AWDLSTM} first map each smart contract as an opcode sequence and then utilize LSTM-based sequence models to make the classification. Compared with \cite{escort} which trains the model in one step, \cite{AWDLSTM} consists of two steps. The first step is training an encoder-decoder architecture to represent the smart contracts. The second step is replacing the decoder layer of AWD-LSTM with some fully-connected layers to make the detection. We use these two methods as baselines because both use bytecode as the input. They all achieve good performance for the vulnerability categories mentioned in their paper. In general, for text classification problems like sentiment analysis, the LSTM-based model can have a good result when the semantic logic is relatively easy. Different from text classification, the logic behind code sequence is more complex. For example, the relation between JUMP and JUMPDEST can be hard to represent and understand in the sequence model. Furthermore, almost the same situation when need to distinguish some complex vulnerability categories. 
All these works\cite{wang2020xg,J2019fuzz,gnn,AWDLSTM,escort} did not adopt good Intermediate Representation like a control flow graph(CFG) to analyze the code execution behavior.


\subsection{Metric Learning}

Classification algorithms in machine-learning-related tasks are well studied. Usually, classification is done by taking the outputs of the last fully connected (FC) layer of the network and feeding into the loss function, i.e., Cross-Entropy, during the learning process. While this process can usually achieve a good result, it ignores the quality of the features generated by the encoder because the loss function only focuses on the final results. Metric learning (or deep metric learning) takes a closer look at features generated by the encoder, especially the feature relation among classes. It tries to optimize such features with some loss functions. It makes the data points belonging to the same class closer and separates the data points belonging to different classes in the latent space.  The optimized representation features help facilitate classification tasks. OAHU\cite{gao2019towards}  propose an  Adaptive-Bound Triplet Loss (ABTL) to effectively utilize the input constraints, and present an Adaptive Hedge Update (AHU) method for updating the model parameters. Compared with general metric learning, i.e., triplet loss, pairwise contrastive (PC) learning conducts a more precise analysis and focuses on the hard samples. It assigns different weights to each pair and adjusts weights dynamically.

We proposed a metric-learning-based framework\cite{vscl} for detecting vulnerabilities in smart contracts and showing the effectiveness and efficiency of our framework. However, this paper extends many sections with more detailed analysis and experimental results. 
Firstly, we compared more related works that utilize machine learning-based methods to detect vulnerabilities and clarify some technical details. This help understands the effectiveness of our developed framework in each stage. We explain the motivation for using the metric learning technique to help the classification. Secondly, we add alternative designs for operation code representation under the current framework and show experimental results accordingly. Thirdly, our proposed framework and its alternatives are evaluated on more datasets constructed in different perspectives with more evaluation metrics to demonstrate the framework's generalized effectiveness. Specifically, in the Majority/Union dataset, the positive label is determined by the voting results of the three tools as mentioned Section~\ref{sec:smart_evaluation}. In the Unknown dataset, positive smart contracts belong to different vulnerable categories in the train and test split. CodeSmell dataset\cite{codesmell2019}, the positives are determined by human experts, and in the SolidFI dataset\cite{ghaleb2020effective}, the positives are constructed by bug injection. In addition, this journal paper brings more detailed information about the internal design of the Ethereum virtual machine since blockchain technology is relatively new. We also illustrate the motivation for using the metric learning technique to help the classification.

\begin{figure}
    \centering
    \fbox{\begin{minipage}{20em}
    00000  PUSH1 0x01 \\
    00002  PUSH1 0x05 \\
    00004  SUB \\
    00005  PUSH1 0x01 \\
    00007  PUSH1 0x05 \\
    00009  ADD 
    \end{minipage}}
    \caption{Addition and subtraction program in EVM bytecode}
    \label{fig:samplecounter}
\end{figure}

\section{Proposed Framework}
\label{sec:smart_arc}
\begin{figure*}
    \centering
    \includegraphics[width=\textwidth]{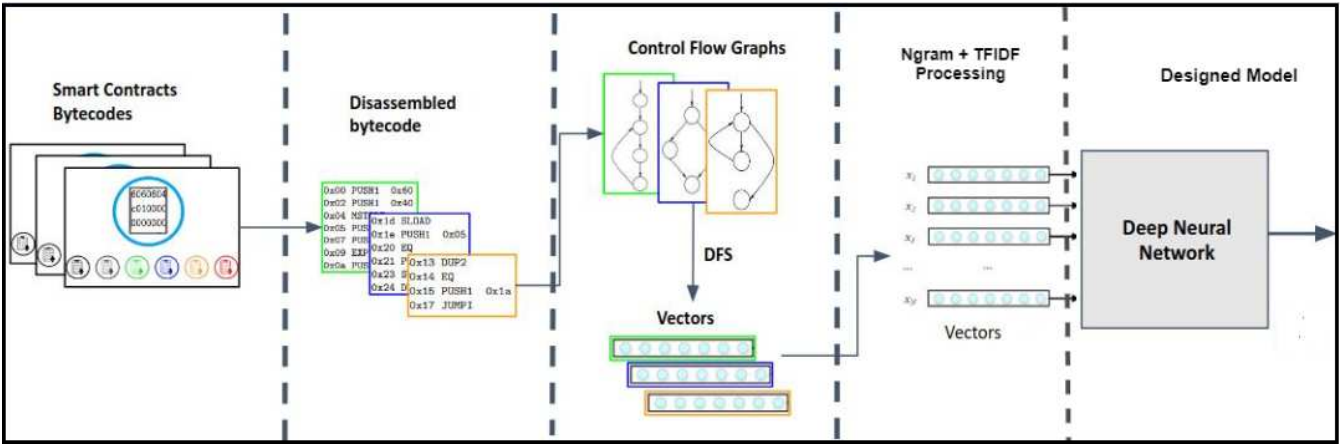}

    \caption{The proposed framework}
    \label{fig:Framework}
\end{figure*}

In this section, we put forward a novel vulnerability detection framework as showing in Figure \ref{fig:Framework}. We first dissemble smart contracts and construct CFGs from which we can infer the internal structure of smart contracts. Then, use encoded vectors to represent CFGs. Finally, we combine all the encoded vectors and feed them into our developed DNN model to make the detection. The following subsections discuss this framework in detail. 

\subsection{Ethereum Virtual Machine Introduction}

The Ethereum virtual machine (EVM) is one of the popular implementations of the smart contract framework. The EVM is a stack-based computer that takes input a sequence of bytecode instructions to complete a task. The EVM has a key-value persistent store, a memory, and a stack of 32-byte values. The EVM bytecode consists of opcodes and each opcode requires a particular set of fees to execute the instruction. The fee varies based on the function of the instruction. Figure \ref{fig:samplecounter} shows a sample EVM bytecode to perform addition and subtraction. In addition, fees are paid to persist data in the smart contract persistent store. Instructions will be executed only after the payment, ensuring that attackers cannot execute their malicious code without wasting their resources.

\subsection{Smart Contract Collection}
In the first step of our proposed framework, we query the Ethereum blockchain repository to get the bytecode of smart contracts. Here, we capture the machine code representation of smart contracts in bytecode form, like the one shows in Figure \ref{fig:bytecode}. 

\begin{figure}
    \centering
    \fbox{\begin{minipage}{25.6em}
    606060405260e060020a6000350463c6c2ea178114601c575b60
    02565b3460025760296004356042565b60408051918252519081
    900360200190f35b605a600283035b600060018211603b575080
    6055565b0190505b919050565b605160018403604256
    \end{minipage}}
    \caption{One Bytecode Example}
    \label{fig:bytecode}
\end{figure}

\subsection{Disassembling}
The second step of the proposed framework is the disassembler, which converts Ethereum bytecode to a sequence of more readable low-level mnemonics\footnote{In programming, a mnemonic is a name assigned to a machine function or an abbreviation for an operation. Each mnemonic represents a low-level machine instruction or opcode in assembly. $add$, $mul$,$lea$,$cmp$,$je$ are examples of mnemonics.}. To do this conversion, the bytecode is scanned, then each instruction is converted to its corresponding mnemonic and incrementing a program counter for each instruction and each inline operand. The output of this step is a series of low-level opcodes and their input arguments.

In our example, the bytecode in Figure \ref{fig:bytecode} is converted to the readable EVM opcodes by the disassembler as shown in Table \ref{Tab:opcode}. 

The opcodes sequence generated here is according to the order of disassembler execution which is not necessarily same as the order of the program execution \cite{ding_dai_yan_zhang_2014}. To make the model understand the program runtime behavior, we need to use the program execution order. Next, we use the CFG to form the program execution order.       

\begingroup
\setlength{\tabcolsep}{10pt} 
\renewcommand{\arraystretch}{1.5} 
\begin{table}[t]
    \centering
    \caption{Disassembling Figure \ref{fig:bytecode} bytecode to opcode}
    \resizebox{\columnwidth}{!}{%
	\begin{tabular}{|l|l| } \hline
		1:  0x0 PUSH1 \hspace*{0.1cm} 0x60 & 12: 0x14 EQ \\ \hline
		2:  0x2 PUSH1 \hspace*{0.1cm} 0x40 & 13: 0x15 PUSH1\hspace*{0.1cm} 0x1c\\ \hline
		3:  0x4 MSTORE & 14: 0x17 *JUMPI \\\hline
		4:  0x5 PUSH1  \hspace*{0.1cm} 0xe0 & 15: 0x18 JUMPDEST\\ \hline
		5:  0x7 PUSH1  \hspace*{0.1cm} 0x02 & 16: 0x19 PUSH1\hspace*{0.1cm} 0x02 \\ \hline
		6:  0x9 EXP  & 17: 0x1B *JUMP \\ \hline
		7:  0xA PUSH1 \hspace*{0.1cm} 0x00 & 18: 0x1C JUMPDEST \\ \hline
		8:  0xC CALLDATALOAD  & 19: 0x1D CALLVALUE \\ \hline
		9:  0xD DIV  & 20: 0x1E PUSH1 \hspace*{0.1cm} 0x02\\ \hline
		10: 0xe PUSH4 \hspace*{0.1cm} 0xc6c2ea17 & 21: 0x20 *JUMPI \\ \hline
		11: 0x13 DUP2  & 22: 0x21 PUSH1\hspace*{0.1cm} 0x29 \\ \hline
		... &  ... \\ \hline
	\end{tabular}
	}
	\label{Tab:opcode}
\end{table}
\endgroup

\subsection{Feature Engineering}
\label{ssec:fea_eng}
This section illustrates the progress of generating feature matrix for all smart contracts. We construct CFGs of smart contracts and then convert to numeric vectors to represent CFGs. The generated feature matrix is used as input of the DNN model in the next section. 
\subsubsection{CFG Construction}

A Control Flow Graph (CFG) is a representation, using graph notation, of control flow or of all paths that might be traversed through the execution of programs. CFGs are mostly used in the static analysis as well as compiler applications, as they can precisely represent the flow inside of a program unit.

\begin{figure}
    \centering
    \includegraphics[scale=0.45]{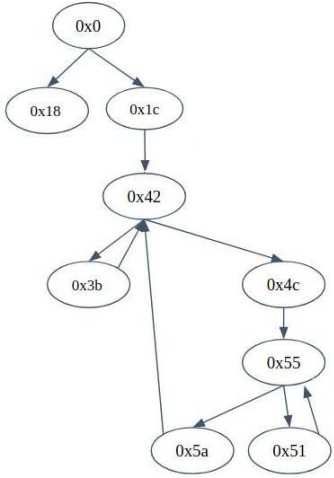}
    \caption{CFG example}
    \label{fig:CGF_example}
\end{figure}

We refer to the approach in \cite{brent2018vandal} to generate the CFG of each smart contract. It represents execution dependency among blocks of codes.  The CFG of an EVM opcode program is initially unknown due to the stack locations, and it needs to be built incrementally and iteratively. To address this problem, we construct the CFG incrementally and we propagate the potential jump addresses' values. In opcode, all JUMP instructions must jump to a \textit{JUMPDEST} instruction. Due to this fact, we can split the disassembled EVM bytecode into basic blocks. In this scenario, the main problem is identifying the JUMPDEST for the JUMP operation since it is not an explicit argument and requires some effort at runtime to pop out of the stack. To overcome this issue, we resolve it by using two phases as described in \cite{brent2018vandal}. After this step, most JUMPDESTs or sets of potential JUMPDESTs for basic blocks have been resolved. Figure \ref{fig:CGF_example} is an CFG example. Each node contains a sequence of instructions. Edges of the node reflect the available paths that the program can take after instructions execution in the node.

\subsubsection{CFG Extraction} \label{sssec:dfs}
We use vectors to represent the CFG of smart contracts. In this step, we adopt depth-first search (DFS) algorithm to traverse the CFG nodes and extract the opcodes in DFS path as shown in Algorithm \ref{fig:DFS}. This algorithm runs on CFGs for all smart contracts and its output is vectors of opcodes where each vector represents the sequence of opcodes for one smart contract. 
Specifically, $opc_{j}$ in $Vector_{i}$ refer to the $jth$ opcode traversed in smart contract ${i}$.
We use this DFS-formed opcode sequence instead of opcode sequence formed by disassembler execution because it better reflects smart contract program run-time scenario \cite{ding_dai_yan_zhang_2014}.  

\begin{equation}
Vector_{i} = \begin{bmatrix}
opc_{1} & opc_{2}& opc_{3} &... &opc_{j} &... & opc_{n}  \\
\end{bmatrix}
\end{equation}

\subsubsection{Construct Feature Matrix} \label{sssec:ngram}
To make the deep learning model catch the patterns in smart contracts, we need to use numeric values to represent the features of smart contract vectors generated in the last step. Here, we adopt n-gram to generate the features for each smart contract and TFIDF to encode each feature. After that, we derive the encoded vectors representing smart contracts and then combine all the encoded vectors to form feature matrix $FM$ as the input of deep learning model.

\paragraph{N-gram}

\begin{figure}
    \centering
    \includegraphics[width=0.28\textwidth]{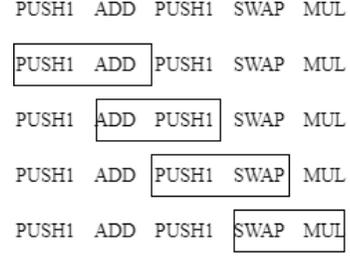}
    \caption{2 gram features generation example}
    \label{fig:2gram}
\end{figure}

We adopt the sliding window approach to obtain n-gram features. A sliding window has a fixed length which equals to ${n}$. When generating the features for each opcode sequence, it will slide from the beginning all the way to the end. Figure~\ref{fig:2gram} shows a process of bigram features generation, where the length of the sliding window is 2. Each distinct group of opcodes within sliding window forms a feature. By adjusting the length of the window, we can obtain grams with different lengths.

 As the length of n-gram increases, more features are generated as more combinations are created among the opcodes. On one hand, this is good for the model as more useful features are fed to make detection. On the other hand, the number of different possible n-gram sequences grows exponentially. It will generate many similar features unavoidably and cause inefficiency especially during the training process. In our case, combining unigram with bigram to form features works best.
\begin{algorithm}[hbtp]
\centering
\caption{Depth First Search on CFG}
\label{fig:DFS}
\begin{algorithmic}[0]
\State \textbf{Depth First Search(control\_flow\_graph)}: \{
\State \hspace*{0.3cm}    \text{visited, stack, opname\_fea = set(), [root], [root.opname]}
\State \hspace*{0.3cm}    while stack: \{
\State \hspace*{0.6cm}        vertex = stack.pop()
\State \hspace*{0.6cm}        if vertex not in visited: \{
\State \hspace*{0.9cm}           visited.add(vertex)
\State \hspace*{0.9cm}           opname\_fea.append(vertex.opname)
\State \hspace*{0.9cm}           stack.extend(graph[vertex] - visited) 
\State \hspace*{0.9cm}          \} 
\State \hspace*{0.6cm}        \}
\State \hspace*{0.3cm}    return opname\_fea 
\State \hspace*{0.3cm} \}
\end{algorithmic}
\end{algorithm}
\paragraph{TFIDF} Term frequency–inverse document frequency (TFIDF) \cite{tfidfdefine} is a numerical statistic that reflects the importance of a word is to a document in a corpus \cite{tfidf}. Here, we weigh each feature generated using n-gram in previous step among all the smart contracts $D$. We see each feature as a term $t$ and each smart contract as a document $d$. The TFIDF value fluctuates proportionally to the number of times a feature appears in the smart contract and it is offset by the number of smart contracts that contains the feature, which helps to adjust to the fact that some features appear more frequently in general. The TFIDF value $tfidf(t,d,D)$ for each feature $t$ in one smart contract $d$ is the product of term frequency $tf(t,d)$ and inverse document frequency $idf(t,D)$. Formulas for $tf(t,d)$, $idf(t,D)$ and $tfidf(t,d,D)$ can be found in \cite{tfidfdefine}.  



After Ngram and TFIDF processing, we have numeric vectors representing smart contracts. Within one smart contract, columns in the vector are the features generated by Ngram and the numeric value in each column is the TFIDF score for that feature. We collect all the numeric vectors to form the feature matrix $FM$. In the feature matrix $FM$, row $i$ is for the encoded vector of smart contract $i$ and feature $f_{i_j}$ is the TFIDF score of feature $j$ in smart contract $i$.

\begin{equation}
FM = \begin{bmatrix}
f_{1,1} & f_{1,2}& f_{1,3} &... & ... & f_{1,features}  \\
   f_{2,1} & f_{2,2}& f_{2,3} &... & ... & f_{2,features}  \\
   . & . & . & . & .& . \\ 
   . & . & . & .  & .& .\\
   f_{i,1} & ... & f_{i,j} & ... &... & f_{i,features}  \\
   . & . & . & . & .& . \\ 
    . & . & . & . & .& . \\ 
\end{bmatrix}
\end{equation}

\begin{figure}[t]

    \includegraphics[scale=0.7]{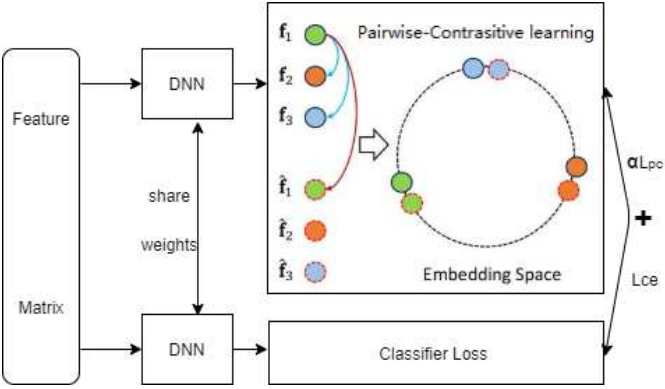}
    \caption{
The framework of our deep learning model. The feature matrix $FM$, which is generated from the Ngram and TFIDF processing, is fed to the DNN. We use the loss function from the pairwise-contrastive learning and the classifier to formulate the model loss function.
 }
    \label{fig:approach}
\end{figure}


\subsection{Deep Learning Model for Classification}
In this step, we apply the deep learning model to our classification task, as shown in Fig. \ref{fig:approach}, our model leverages deep neural networks (DNNs) to serve as a classifier model. Specifically, besides the typical used softmax/cross-entropy loss in DNNs, we introduce a metric learning function on the last hidden layer output, which works in the embedding space as a regularization term to compress the intra-class embedding distance (positive) and separate the embedding distance from inter-class (negative). In contrast to existing DNNs that only apply the cross entropy loss function, our approach learns a more discriminated model \cite{meyer2018deep,wang2019co} to execute the classification task.

\textbf{Metric Learning}
We sample different vector pairs from the feature matrix $FM$ and form the pair-set in every mini-batch $\mathcal{U}$, specifically. We first randomly generate $m$ vector pairs from the input matrix $FM$, then generate the pair-constraint $C = \lbrace (x_i,x_j,I), x_i,x_j\in FM\rbrace_m$, in every $\mathcal{U}$. Here the $I \in \lbrace -1,1 \rbrace^{m}$ is the indicator that describes the similarity of corresponding $(x_i,x_j)$ (\textit{i.e.} $I_{i} = 1$ if $x_i$ and $x_j$ have the same class, otherwise $I_{i} = -1$). We define the metric learning model with parameter $\theta$ as $\phi(\theta)$. The similar measurement $S(x_{i},x_{j}) = S(\phi(x_i;\theta), \phi(x_j;\theta))$ represents the similarity score (\textit{e.g.} calculated from the cosine similarity function) of the pair $(x_i, x_j)$ in the embedding space. 

\textbf{Similarity Analysis on Vector Pairs:}
In terms of similarity $S$ for the input vector pair, the derivative of the loss function $L$ based on similarity metrics $L(S)$ with respect to the model parameters $\theta$ could be described as: $\frac{\partial L}{\partial \theta} = \sum_{i}\sum_{j} \frac{\partial L(S)}{\partial S_{ij}} \frac{\partial S_{ij}}{\partial \theta}$. This indicates the contrastive loss can be reformulated to a new form of pairwise weighting $w_{ij} = \frac{\partial L(S)}{\partial S_{ij}}$. For the optimization on the loss function $L(S)$, the drawback of existing triplet/contrastive loss is that all the selected pairs $(x_i, x_j)$ are assigned with equal weight \cite{duan2018deep,zagoruyko2015}. However, considering the ``hard negative" vectors (the feature vectors that are closer to various classes' boundary \cite{duan2018deep}), it is obviously that such vectors need more attention by assigning with larger weights. Here, we prefer to select more informative pairs (\textit{e.g.}, anchor with hard negative ones) and make the majority contribution to the learning process. In our work, we consider a novel pair weighting strategy to address the mining of such informative negative/positive pairs. 

\textbf{Weighting Strategy for Pairwise:}
Our approach implements a novel strategy for pairwise weighting. We first calculate the relative similarity (for the positive/negative pairs) based on the same anchor vector. Then we sample the significant ones from such pairs. We define the similarity between positive vector pairs as $S_{ij}^{+} = (x_{i},x_{j})$ (if $x_{i}$ and $x_{j}$ belong to the same category), and $S_{ij}^{-} = (x_{i},x_{j})$ (if $x_{i}$ and $x_{j}$ come from different classes) as the similarity from negative pairwise. Given the anchor vector $x_i$, a positive pair $(x_i, x_j)$ is sampled by comparing it to a negative vector that has the largest similarity with $x_i$. This can be described as: $S_{ij}^{+} < \max_{y_i \neq y_k} (S_{ik}) + \eta$. On the other hand, a negative pair of vectors is compared to the positive vector that has the lowest similarity with $x_i$. It hence can be summarised as: $S_{ij}^{-} > \max_{y_i = y_k} (S_{ik}) - \eta$, where $\eta$ is a given margin value. Next, in order to optimize the weight for various pairs, we apply a soft weighting strategy \cite{wang2019multi} to make the selected pairs more effectively. For a sampled pair $(x_i,x_j)$, the weight of pairs ($w_{ij}^{+}, w_{ij}^{-}$) can be computed as:
\begin{equation}
\left\{
\begin{aligned}
w_{ij}^{+}= \frac{\exp{(\lambda_{1}(\omega - S(x_i,x_j)))}}{1 + \sum_{k \in C^{+}} \exp{(\lambda_{1}(\omega - S(x_i,x_k)))}} \text{,  \qquad if } y_{i} = y_{j} \\
w_{ij}^{-}= \frac{\exp{(\lambda_{2}(S(x_i,x_j) - \omega))}}{1 + \sum_{k \in C^{-}} \exp{(\lambda_{2}(S(x_i,x_k) - \omega))}} \text{,  \qquad if } y_{i} \neq y_{j}
\end{aligned}
\right.
\label{eq:weight}
\end{equation}

where $\lambda_{1}, \lambda_{2}$ and $\omega$ are hyper parameters in the Binomial deviance \cite{yi2014deep},  $C^{+}$ and $C^{-}$ denote the positive and negative pair set in $C$. Our weighting strategy updates the weights of different pairs during the training step in a dynamic manner.

Finally, we integrate pair mining and our weighting scheme into a single framework, and provide a new vector pair based loss function to describe the relationship, namely pairwise-contrastive (PC) loss:
\begin{equation}
\begin{split}
&\ell_{PC}=\sum_{i=1}^{m} \Big \{\frac{1}{m^{+}}\sum_{y_{i}=y_{j}}g(\lambda_{1}[\omega - S(x_i,x_j)])\\
&+ \frac{1}{m^{-}}\sum_{y_{i}\neq y_{j}}g(\lambda_{2}[S(x_i,x_j)-\omega])  \Big \}
\label{eq:mc_loss}
\end{split}
\end{equation}

where $g(x) = \log(1+\exp(x))$ is a generalized logistic loss function and it is typically used as approximation of hinge loss. Our PC loss can be minimized with stochastic gradient descent (SGD) based optimization approach, in which we combine with the proposed iterative sampling and weighting strategy during every iteration step. The loss minimized by our deep learning model in Fig. \ref{fig:approach} is describe below: 
\begin{equation}
\begin{split}
\ell = \ell_{CE} + \alpha \ell_{PC}
\label{eq:combine_loss}
\end{split}
\end{equation}

where $\ell_{CE}$ is the cross entropy loss from the DNN classifier, $\alpha$ is a fixed scalar hyperparameter that denotes the relative weight of different loss functions. In section \ref{sec:Ablation}, we show the advantage of combining the two loss functions by comparing the learned feature embedding and their performances accordingly.
 
\subsection{Alternatives for Feature Matrix}
\label{sec:AlternativesFeature Matrix}
After the step of Section \ref{sssec:dfs} , we form a opcode sequence for each smart contract. And in the following Section \ref{sssec:ngram}, we combine Unigram and Bigram to form features and then use TFIDF to weigh each feature. All the encoded smart contracts together form a feature matrix and then feed into the classification model. Ngram+TFIDF in section \ref{sssec:ngram} is best suitable for our deep-learning model and current Ethereum dataset, while some other alternative methods to form the feature matrix can also have relatively good performance.  These methods include  Integer Encoding, Unigram + TFIDF, Long Former \cite{beltagy2020longformer}, Doc2Vec \cite{le2014distributed}. In addition, these alternative methods can be useful for other tasks besides vulnerability detection or smart contract datasets in other blockchains. Hence, we illustrate the implementation details about other possible encoding methods to form feature matrix $FM$ and then compare the performance in the following experiment section. 

\textbf{Integer Encoding} 
is a basic encoding approach where each unique opcode is mapped to an integer. For example, the previous opcode sequence in Figure~\ref{fig:2gram}, if we use 1 to represent $ADD$, 2 to represent $PUSH1$, 3 to represent $SWAP$ and 4 to represent $MUL$, the encoded sequence should be $2,1,2,3,4$. Due to these integers being randomly given for each opcode, Integer Encoding ignores any possible existing ordinal relationships between each opcode and leave these for the model to learn. To get the fixed length feature vector, we consider the maximum number $l$ of this feature vector. It means if the length of the opcode vector is less than $l$, we add zeros (padding) to the end of it. Otherwise, we only consider the first $l$ entries in the opcode vector $Vector_{i}$ for each smart contract ${i}$. Here we set $l$=2048 which has the best performance among others.

\textbf{Unigram + TFIDF} 
Like our approach, Unigram + TFIDF first forms fixed numbers of features using Unigram and then utilizes TFIDF to score these features. The only difference is it only uses Unigram to form features instead of both Unigram and Bigram. In other words, for Unigram, the sliding window length in Figure~\ref{fig:2gram} is equal to 1. So, this method considers each opcode as a feature and then use TFIDF to score each feature.

\textbf{Longformer}  is a transformer model which can analyze long sequences. We use the pre-trained Longformer model: longformer-base-4096 as the encoder. The reason we choose Longformer but not BERT \cite{devlin2019bert} is BERT is unable to process sequences longer than 512 due to its self-attention operation, which scales quadratically with the sequence length. The Longformer has an attention mechanism that scales linearly with sequence length, making it easy to process documents of thousands of tokens or longer. For our smart contracts dataset, most opcodes sequences are long than 512.

\textbf{Doc2Vec} is an extension to Word2vec\cite{mikolov2013efficient}. Word2Vec learns to project words into a latent space whereas Doc2Vec aims at learning how to project a document into a latent space. Here we see each smart contract as one documnet and use gensim 4.0.0 doc2vec \cite{rehurek_lrec} as the encoding model. The output of each smart contract is a fixed size vector. We set $vector\_size=500$, $window=5$, $min\_count=1$ when training.  

\section{Evaluation}
\label{sec:smart_evaluation}
In this section, we describe the benchmark and present the experimental results.


\subsection{Benchmarks}
\subsubsection{Collected Smart contracts datasets}

We retrieved the bytecode of smart contracts deployed on the Ethereum blockchain and removed duplicated contracts. In total, there are 205,848 unique smart contract bytecodes. Ethereum contracts vary enormously from simple to very complex. The number of the instructions in Ethereum contracts varies from 8 to 13,050, with an average of 1,545.27 and a median of 1,213. The number of distinct instructions in the contracts varies from 7 to 57. 

In this project, we use EVM bytecode instead of smart contract source code since most smart contracts’ source codes are not available on public repositories.
\begin{figure}[t]
    \centering
    \includegraphics[scale=0.6]{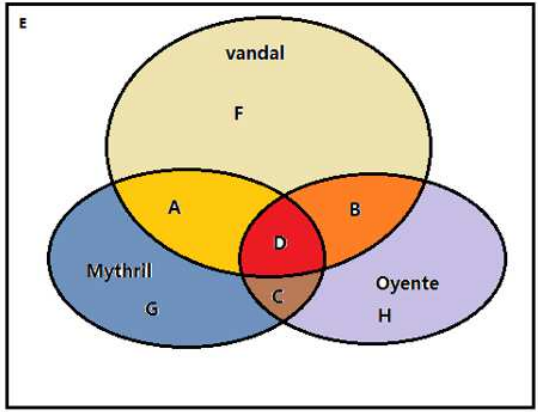}
    \caption{The number of smart contracts in each regions are Rectangle:160012, A:28074, B:10194, C:907, D:3789,  E:32735, F:73699, G:6309, H:4308  }
    \label{fig:label_diagram}
\end{figure}
 
 \textbf{Label determination} Smart contracts we collected are from the real-world, and many of them are actually in use every day. We select to use three tools - Oyente, Mythril, and Vandal as the reference to determine the label. They are the mainstream tools to deal with smart contract bytecodes analysis, and many existing  approaches\cite{kalra2018zeus,wang2020xg,J2019fuzz} use them as the benchmark. Here, we discard those contracts if one of these three tools could not get the label in 60 seconds.
 
To determine a generally reliable label for every smart contract, we take the testing results from all three tools into consideration.
However, there are two main difficulties in determining labels. (1) They are hard to agree on the vulnerability category level. Some tools apply conservative rules, while others apply strict rules, especially for reentrancy vulnerability, Oyente, Vandal, and Mythril claim 2784, 104757, 78 smart contracts individually. (2) These tools define vulnerability categories with different names, and some categories are overlapped when defined by different tools. To address this issue, we convert all smart contracts containing vulnerabilities as one class, set their labels as positive, and set the label of the benign class as negative. And then, by using the labels from these three tools, we create two datasets namely \textbf{Majority dataset} and \textbf{Union dataset}. For the Majority dataset, in each smart contract, if there is a vulnerability in two of these three tools, we set the label as positive and negative otherwise. For the Union dataset, if there is a vulnerability in any of these three tools in each smart contract, we set the label as positive and negative otherwise. These two datasets are used to balance the strict and conservative rules in these tools and show our proposed framework has a good performance in both settings. 
 
Figure \ref{fig:label_diagram} is plotted to better illustrate the smart contracts labels' statistics. The rectangle area represents the smart contracts we collected in total, and the area within each circle represents the smart contracts found with vulnerabilities by each tool. Each circle represents a tools and there are three tools in total. The overlapped area among the circles indicates multiple tools found the smart contracts vulnerable. For example, Region A represents the smart contracts found vulnerabilities by Vandal and Mythril, and Region D means the smart contracts found vulnerabilities by all these three tools. The white area within the rectangle indicates all three tools found the smart contracts without any vulnerabilities. The number of smart contracts in each region is listed in Figure \ref{fig:label_diagram}. For the Majority dataset, we set the smart contracts within the region A, B, C, D as positive while others as negative. For the Union dataset, we set the smart contracts within the three circles as positive while the rest are negative. In the Majority dataset,  117,048 smart contracts belong to the benign (negative) class, and 42,964 smart contracts belong to the vulnerable (positive) class. In the Union dataset, 32,735 smart contracts belong to the benign (negative) class, and 127,277 smart contracts belong to the vulnerable (positive) class.
\subsubsection{Experts Manually Determined Dataset}Codesmell\cite{codesmell2019} datasets contain vulnerable smart contracts with 20 vulnerability categories. The smart contracts were collected from posts of Ethereum StackExchange, as well as real-world smart contracts. Human experts from 32 countries manually determine the labels.

\subsubsection{Injected Vulnerabilities Dataset}SolidiFI\cite{ghaleb2020effective} dataset contains buggy contracts injected by 9369 bugs from 7 different bug types: reentrancy, timestamp dependency, unhandled exceptions, unchecked send, TOD, integer overflow/underflow, and use of tx.origin. The bugs have been injected in the contracts using the SolidiFI framework. In SolidiFI dataset, we compile source code to provide bytecode using the compiler version as suggested.

\subsection{Neural network configuration}
\label{sec:network_configuration}
Figure \ref{fig:approach} highlights the structure of our model. The input is the feature matrix $FM$ which is constructed after feature engineering step. DNN model serves as a feature selector for the Pairwise-Contrastive Learning function. Those two DNN models which share the weights in Figure \ref{fig:approach} have the same configuration. It has 5 hidden layers and activation function we use is Relu. For the Pairwise-Contrastive Learning function part, we set $\lambda_{1}$ = 2, $\lambda_{2}$ = 40, $\omega$ = 0.5, $S$ to be the cosine similarity in Eq.~\ref{eq:weight} and ~\ref{eq:mc_loss}. In Eq.~\ref{eq:combine_loss}, $\alpha$ is used for weight balancing between loss functions, which is initialized to 0.8. Total number of epochs is $300$ and learning rate $\eta$=0.02. The loss function is optimized by stochastic gradient descent(SGD) with momentum.

\subsection{Baseline models and configuration}
We use Logistic Regression, k-nearest neighbors(KNN), Random Forest, Adaboosting, another metric learning-based DNN(OAHU) \cite{gao2019towards}, and two LSTM-based models (ESCORT \cite{escort} and AWDLSTM \cite{AWDLSTM} ) as baselines. In Logistic Regression, max number of iterations is 2000, and use $l2$ penalty. In  k-nearest neighbors(KNN), $K=3$ and using the minkowsk metric. In Random Forest, we set the number of estimators to be 80. In OAHU, we set $\beta=0.99$ (decay factor), $
hidden\_layer=5$, $S_{hidden}=4911$ (number of units in each hidden layer), $s=0.1$ (smooth factor), $\eta=0.3$ (learning rate), $S_{emb}=4911$ (dimensionality of learned metric embedding), $\tau=0.5$ (effective radius), iteration num=$10^4$ and ADAM as optimizer. In ESCORT \cite{escort} and AWDLSTM \cite{AWDLSTM}, we follow default settings in both smart contract encoding and parameters selection.

\subsection{Experimentation}

\subsubsection{Experimental Criteria}

All parts of the experiment were conducted on an Intel machine having Core-i9-7920X 2.90GHz CPU with 125 GB of RAM and GeForce RTX 2080 Ti GPU, running a standard Ubuntu version 18.04.4 LTS. 

In the Majority and Union dataset, we randomly split data into training and test sets with the ratio of 80\% and 20\%, respectively. In the CodeSmell and SolidiFI dataset, for both training and test sets, the number of benign smart contracts is 1000, and the number of malicious smart contracts is 200. The benign smart contracts are randomly chosen from the datasets we collected. Hyperparameters 
were selected by ten-fold cross validation (CV) procedure.
We use accuracy, precision (P), recall (R), F1-score and weighted F1-score as the evaluation metrics. 
Accuracy is able to offer a general performance overview of each model.
In a vulnerability detection scenario, it is required to focus more on models' non-benign class performance especially when the data is imbalanced. We use precision (P), recall (R), F1-score to give a view from this perspective.
The precision is defined as $P = TP/(TP+FP)$ and recall as $ R = TP/(TP+FN)$, where $TP$, $FP$ and $FN$ denote the number of true positives, false positives and false negatives, respectively. F1-score is commonly defined as the harmonic mean of precision and recall: $F1 = 2$ $\times$ $P$ $\times$ $R/(P+R)$. The weighted F1-score is the averaged F1-scores for all classes. Hence, it measures the F1 for all classes. 
 
\subsubsection{Experimental Results}
We compare the performance of the proposed methods with seven baseline models. Logistic Regression, KNN (K=3), Random Forest (RF),  Adaboosting, and one metric learning-based model OAHU \cite{gao2019towards} are selected to conduct model classification evaluation. We use the smart contract representation method in \ref{ssec:fea_eng} to form the input feature matrix. For baselines used for whole framework comparison(ESCORT \cite{escort}, and AWDLSTM \cite{AWDLSTM}), we follow their original experiment setting.

The performance for the 4 datasets (Majority,  Union, CodeSmell and SolidiFI) are shown in Table \ref{Tab:majority}, \ref{Tab:union}, \ref{Tab:codesmell} and \ref{Tab:solidifiy} accordingly. Across all the datasets, all methods have relatively high accuracy, even for relatively simple models like KNN and Logistic Regression, which is above 80\%. It reflects that the smart contract representation method is reasonable enabling models to understand the smart contract. On the other hand, our model outperforms the rest baseline methods. 
It demonstrates Pairwise-Contrastive Learning approach separates instances from different classes well and makes instances of the same class closer in the new latent feature space. After ours, ensemble models (Random Forest, adaboosting) generally perform better than the other metric learning model OAHU\cite{gao2019towards} and two LSTM-based models (ESCORT \cite{escort} and AWDLSTM \cite{AWDLSTM}). The traditional machine learning methods (Logistic Regression, KNN (k=3)) have the worst performance. The performance of all models in the Union, CodeSmell, and SolidiFI datasets are better than the Majority dataset.

We also compare True Positive Rate (TPR) vs. False Negative Rates (FNR) and True Negative Rate (TNR) vs. False Positive Rate (FPR) in Figure \ref{fig:tprfnr123}. In general, higher TPR and TNR indicate better model performance. With the majority dataset, all models generally perform well on TNR vs. FPR. As for TPR vs. FNR, the TPR of our model is 68\%, whereas the TPR of others ranges from 51\% to 67 \%. With the union dataset, for TNR vs. FPR, the TNR of our model is 95\%, which is 4\% higher than the second high, i.e., Adaboosting. All models generally perform well for TPR vs. FNR. The performances in the CodeSmell and SolidiFI datasets follow the same pattern as the Union dataset.
 
Overall, the performance results from 4 datasets across nine total evaluation metrics show that our model has a consistently good performance in detecting the vulnerabilities of the smart contracts whenever defining vulnerabilities from different perspectives.
\begingroup
\setlength{\tabcolsep}{2pt} 
\renewcommand{\arraystretch}{1.2} 
\begin{table}
	\centering
	\caption{Majority Dataset Performance}
	\resizebox{\columnwidth}{!}{%
	\begin{tabular}{|c|c|c|c|c|c| } \hline
		\textbf{Model}& \textbf{Accuracy}& \textbf{Precision} & \textbf{Recall} & \textbf{F1 score } & \textbf{Weighted F1 score} \\ \hline
		Logistic Regression &  0.86 & \textbf{0.88} &  0.57 &  0.69 & 0.85\\ \hline
		KNN (k=3) & 0.87 & 0.81 & 0.67 & 0.73 & 0.86\\ \hline
	    Random Forest  & 0.88 &  0.87 & 0.64 & 0.74 & 0.87 \\ \hline
	    Adaboosting  &  0.87 &  0.86  &  0.60 &  0.71 & 0.86 \\ \hline
	    OAHU \cite{gao2019towards}  &  0.84 &  0.84  &  0.51 &  0.63 & 0.83 \\ \hline
	    ESCORT \cite{escort}  &  0.85 &  0.74  &  0.57 &  0.64 & 0.84 \\ \hline
	    AWDLSTM \cite{AWDLSTM}  &  0.84 &  0.73  &  0.55 &  0.63 & 0.83 \\ \hline
		\textbf{Ours}  & \textbf{0.89} & 0.87 & \textbf{0.68} &  \textbf{0.76} & \textbf{0.88} \\ \hline

	\end{tabular}
	}
	\label{Tab:majority}
\end{table}

\begingroup
\setlength{\tabcolsep}{2pt} 
\renewcommand{\arraystretch}{1.2} 
\begin{table}
	\centering
	\caption{Union Dataset Performance}
	\resizebox{\columnwidth}{!}{%
	\begin{tabular}{|c|c|c|c|c|c| } \hline
		\textbf{Model}& \textbf{Accuracy}& \textbf{Precision} & \textbf{Recall} & \textbf{F1 score } & \textbf{Weighted F1 score} \\ \hline
		Logistic Regression &  0.92 &  0.95 & \textbf{0.95} &  0.95 & 0.92\\ \hline
		KNN (k=3) & 0.91 & 0.95 & 0.94 & 0.94 & 0.91\\ \hline
	    Random Forest  & 0.94 &  0.97 & \textbf{0.95} & 0.96 & 0.94 \\ \hline
	    Adaboosting  & 0.94 & 0.98 & 0.94 &  0.96 & 0.94 \\ \hline
	    OAHU \cite{gao2019towards}  &  0.88 &  0.92  &  0.93 &  0.93 & 0.88 \\ \hline
	    ESCORT \cite{escort}  &  0.92 &  0.94  &  0.95 &  0.95 & 0.92 \\ \hline
	    AWDLSTM \cite{AWDLSTM}  &  0.91 &  0.95  &  0.94 &  0.94 & 0.91 \\ \hline
		\textbf{Ours}  & \textbf{0.95} & \textbf{0.99} & \textbf{0.95} &  \textbf{0.97} & \textbf{0.95} \\ \hline

	\end{tabular}
	}
	\label{Tab:union}
\end{table}

\begingroup
\setlength{\tabcolsep}{2pt} 
\renewcommand{\arraystretch}{1.2} 
\begin{table}
	\centering
	\caption{CodeSmell Dataset Performance}
	\resizebox{\columnwidth}{!}{%
	\begin{tabular}{|c|c|c|c|c|c| } \hline
		\textbf{Model}& \textbf{Accuracy}& \textbf{Precision} & \textbf{Recall} & \textbf{F1 score } & \textbf{Weighted F1 score} \\ \hline
		Logistic Regression &  0.92 &  \textbf{0.90} & 0.13 &  0.23 & 0.89\\ \hline
		KNN (k=3) & 0.93 & 0.61 & 0.58 & 0.59 & 0.93\\ \hline
	    Random Forest  & 0.95 &  0.78 & 0.60 & 0.68 & 0.94 \\ \hline
	    Adaboosting  & 0.95 & 0.81 & 0.61 &  0.70 & \textbf{0.95} \\ \hline
	    OAHU \cite{gao2019towards}  &  0.95 &  0.67  &  \textbf{0.63} &  0.65 & 0.94 \\ \hline
	    ESCORT \cite{escort}  &  0.93 &  0.64  &  0.60 &  0.62 & 0.93 \\ \hline
	    AWDLSTM \cite{AWDLSTM}  &  0.92 &  0.58  &  0.57 &  0.57 & 0.92 \\ \hline
		\textbf{Ours}  & \textbf{0.96} & 0.85 & \textbf{0.63} &  \textbf{0.72} & \textbf{0.95} \\ \hline

	\end{tabular}
	}
	\label{Tab:codesmell}
\end{table}

\begingroup
\setlength{\tabcolsep}{2pt} 
\renewcommand{\arraystretch}{1.2} 
\begin{table}
	\centering
	\caption{SolidiFI Dataset Performance}
	\resizebox{\columnwidth}{!}{%
	\begin{tabular}{|c|c|c|c|c|c| } \hline
		\textbf{Model}& \textbf{Accuracy}& \textbf{Precision} & \textbf{Recall} & \textbf{F1 score } & \textbf{Weighted F1 score} \\ \hline
		Logistic Regression &  0.95 &  \textbf{0.98} & 0.90 &  0.94 & 0.95\\ \hline
		KNN (k=3) & 0.95 & \textbf{0.98} & 0.90 & 0.93 & 0.95\\ \hline
	    Random Forest  & 0.97 &  0.96 & 0.96 & 0.96 & 0.97 \\ \hline
	    Adaboosting  & 0.97 & 0.96 & 0.96 &  0.96 & 0.97 \\ \hline
	    OAHU \cite{gao2019towards}  &  0.96 &  0.95  &  0.94 &  0.95 & 0.96 \\ \hline
	    ESCORT \cite{escort}  &  0.96 &  0.95  &  0.95 &  0.95 & 0.96 \\ \hline
	    AWDLSTM \cite{AWDLSTM}  &  0.96 &  0.95  &  0.95 &  0.95 & 0.96 \\ \hline
		\textbf{Ours}  & \textbf{0.98} & 0.97 & \textbf{0.97} &  \textbf{0.97} & \textbf{0.98} \\ \hline

	\end{tabular}
	}
	\label{Tab:solidifiy}
\end{table}

\begin{figure*}
        \centering
        \begin{subfigure}[b]{0.46\textwidth}
            \includegraphics[width=\textwidth]{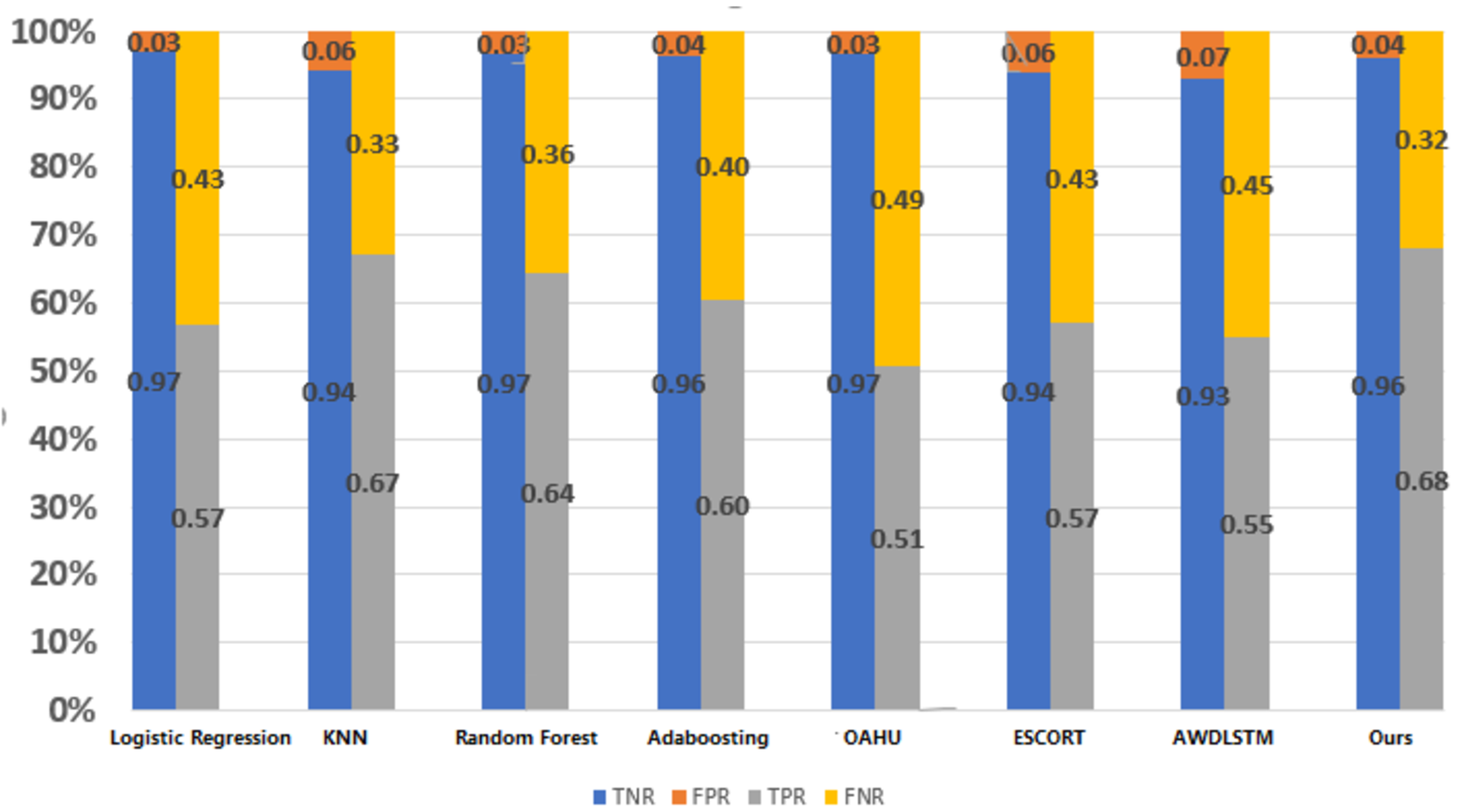}
            \caption{Majority Dataset}
        \end{subfigure}
        \begin{subfigure}[b]{0.46\textwidth}
            \includegraphics[width=\textwidth]{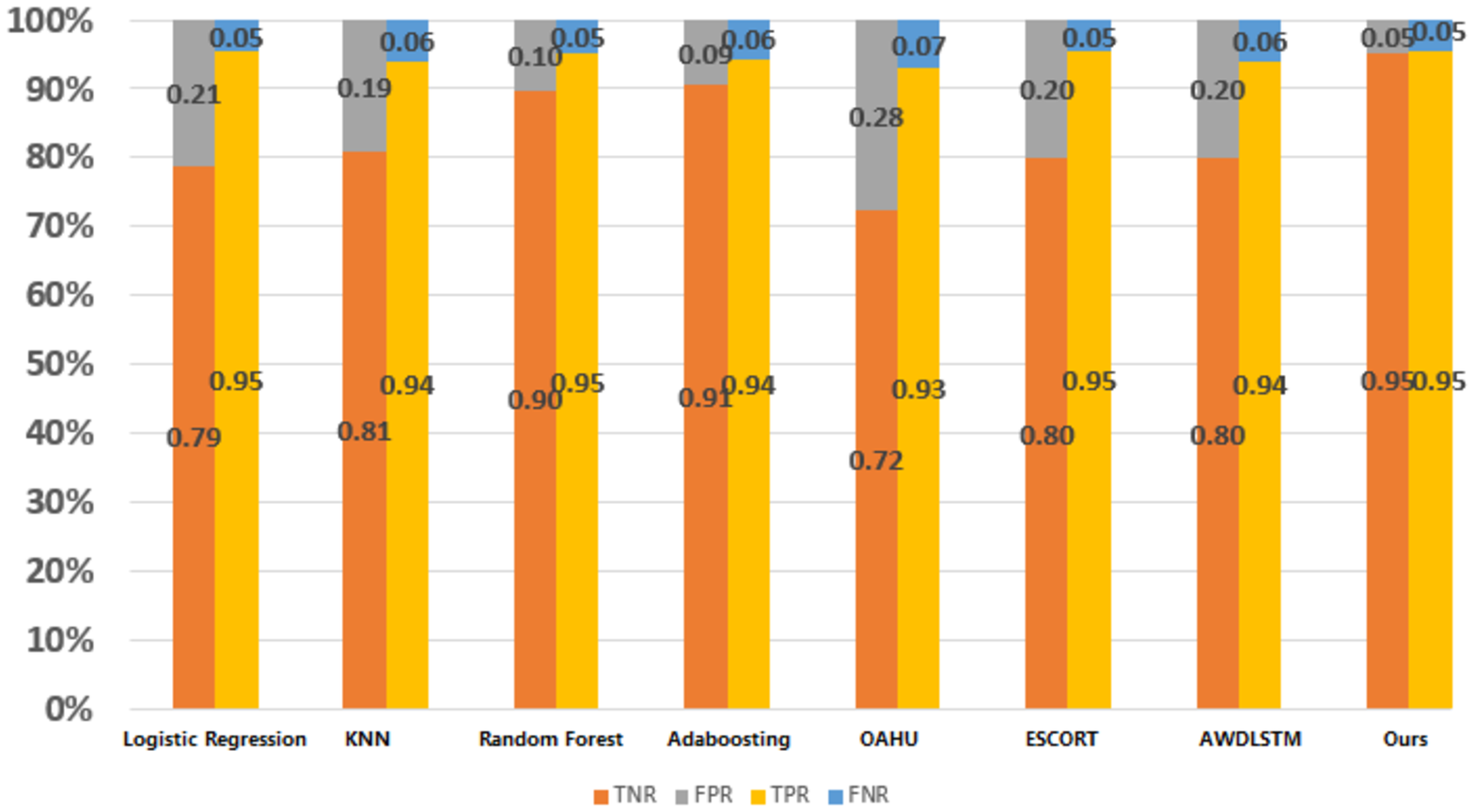}
            \caption{Union Dataset}
        \end{subfigure}
        \begin{subfigure}[b]{0.46\textwidth}
            \includegraphics[width=\textwidth]{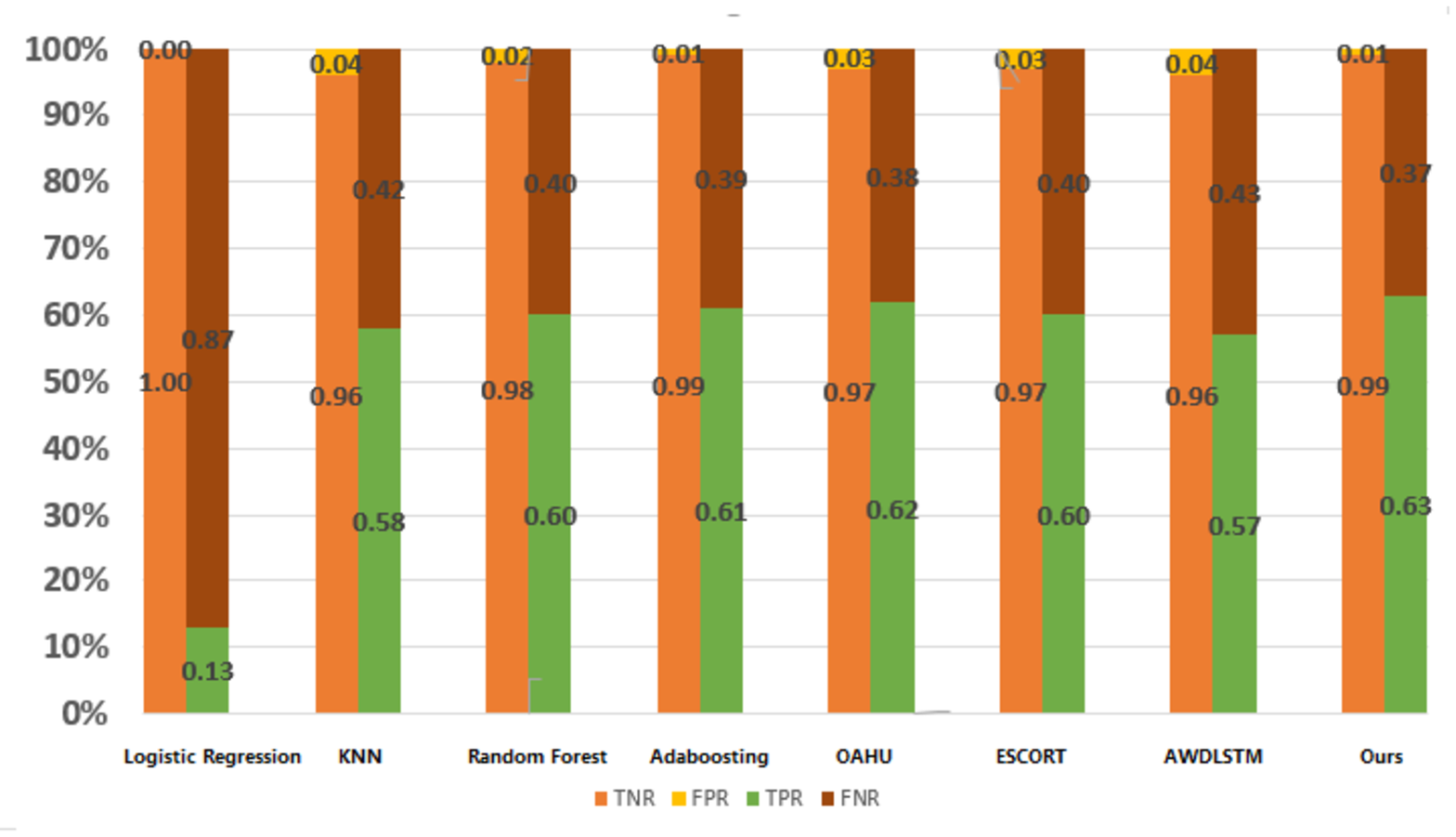}
            \caption{CodeSmell Dataset}
        \end{subfigure}
        \begin{subfigure}[b]{0.46\textwidth}
            \includegraphics[width=\textwidth]{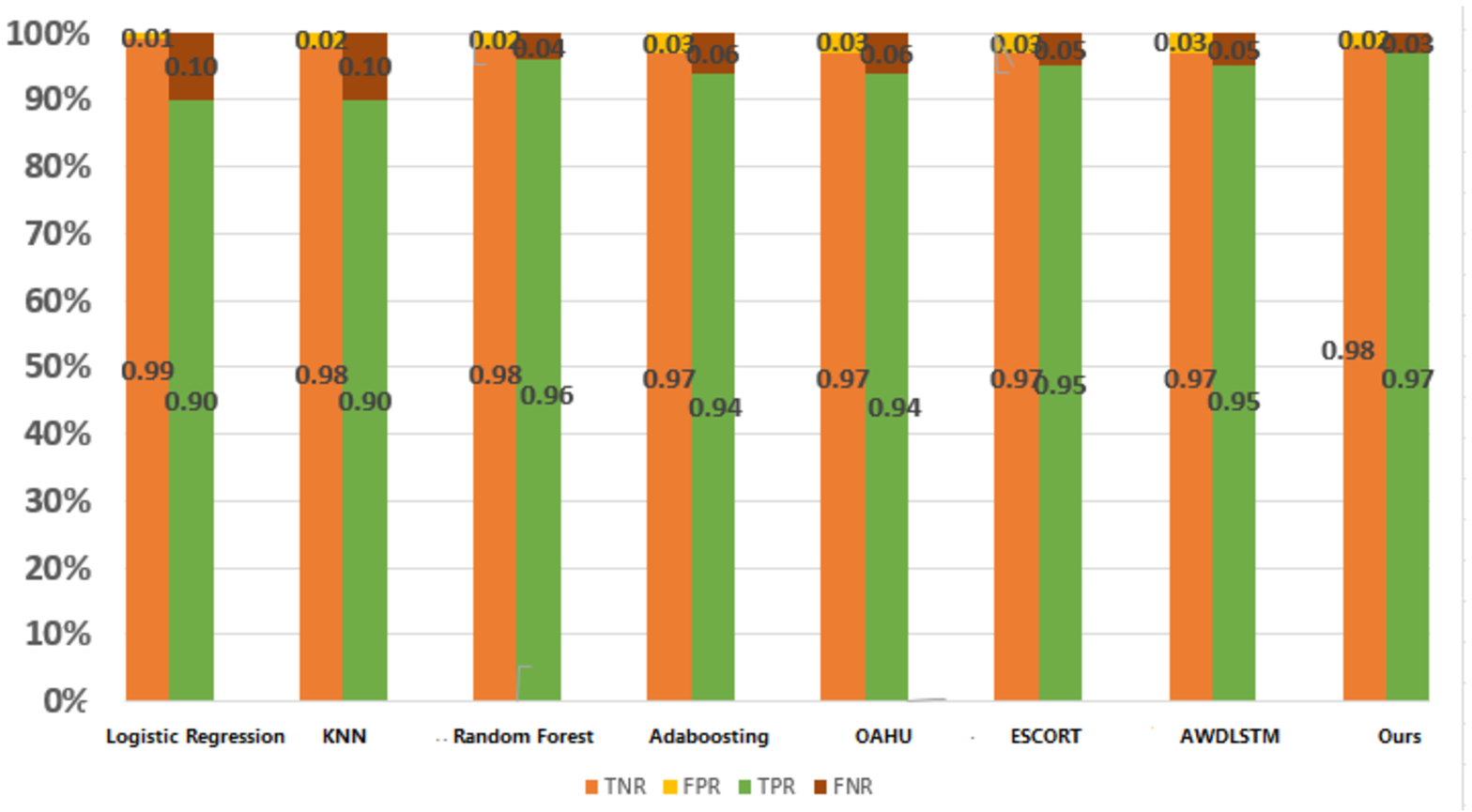}
            \caption{SolidiFI Dataset}
        \end{subfigure}
        \vspace{-3mm}
        \caption{TPR, FNR, TNR and FPR Comparison}
        \label{fig:tprfnr123}
        \vspace{-3mm}
\end{figure*}

\begingroup
\setlength{\tabcolsep}{2pt} 
\renewcommand{\arraystretch}{1.2} 
\begin{table}
	\centering
	\caption{Different Potential Encoding Methods Comparison}
	\resizebox{\columnwidth}{!}{%
	\begin{tabular}{|c|c|c|c|c|c| } \hline
		\textbf{Model}& \textbf{Accuracy}& \textbf{Precision} & \textbf{Recall} & \textbf{F1 score } & \textbf{Weighted F1 score} \\ \hline
		Integer Encoding &  0.85 &  0.82 & 0.57 &  0.67 & 0.84\\ \hline
		Unigram+TFIDF & 0.86 & 0.80 & 0.63 & 0.70 & 0.86\\ \hline
	    Long Former \cite{beltagy2020longformer}  & 0.85 &  0.80 & 0.61 & 0.69 & 0.85 \\ \hline
	    DOC2VEC \cite{le2014distributed}  & 0.86 & 0.79 & 0.67 &  0.73 & 0.85 \\ \hline
		\textbf{Ours}  & \textbf{0.89} & \textbf{0.87} & \textbf{0.68} &  \textbf{0.76} & \textbf{0.88} \\ \hline

	\end{tabular}
	}
	\label{Tab:encoding}
\end{table}

 \begin{figure*}[t]
\centering
\includegraphics[width=1.6\columnwidth]{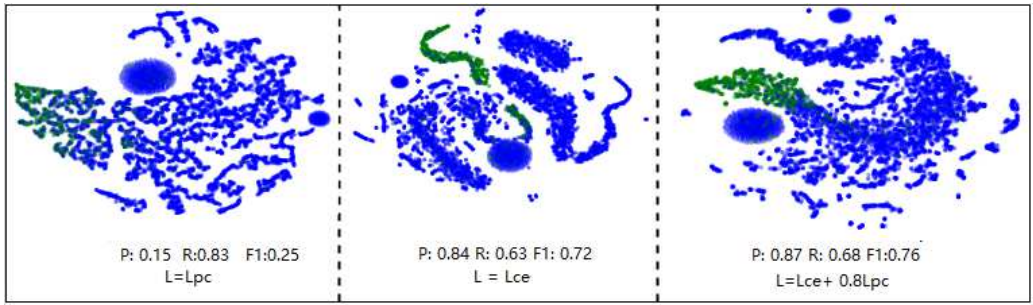}
\caption{
Visualization of learned feature embedding when choosing different loss function. Blue dots are for benign class and green dots are for vulnerable class.
 }
\label{fig:visu}
\end{figure*}
\begin{figure*}
        \centering
        \begin{subfigure}[b]{0.48\textwidth}
            \includegraphics[width=\textwidth]{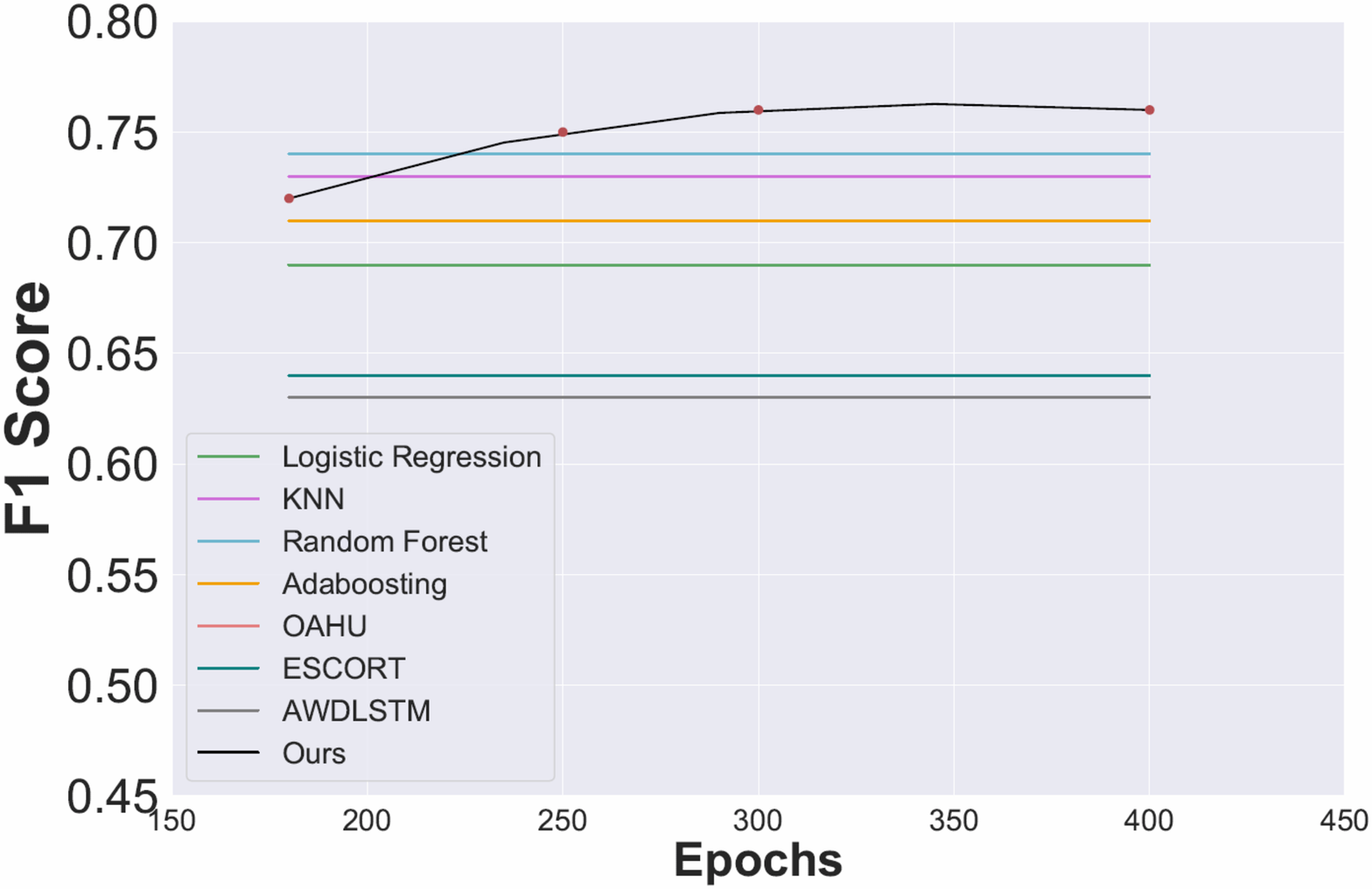}
        \end{subfigure}
        \begin{subfigure}[b]{0.48\textwidth}
            \includegraphics[width=\textwidth]{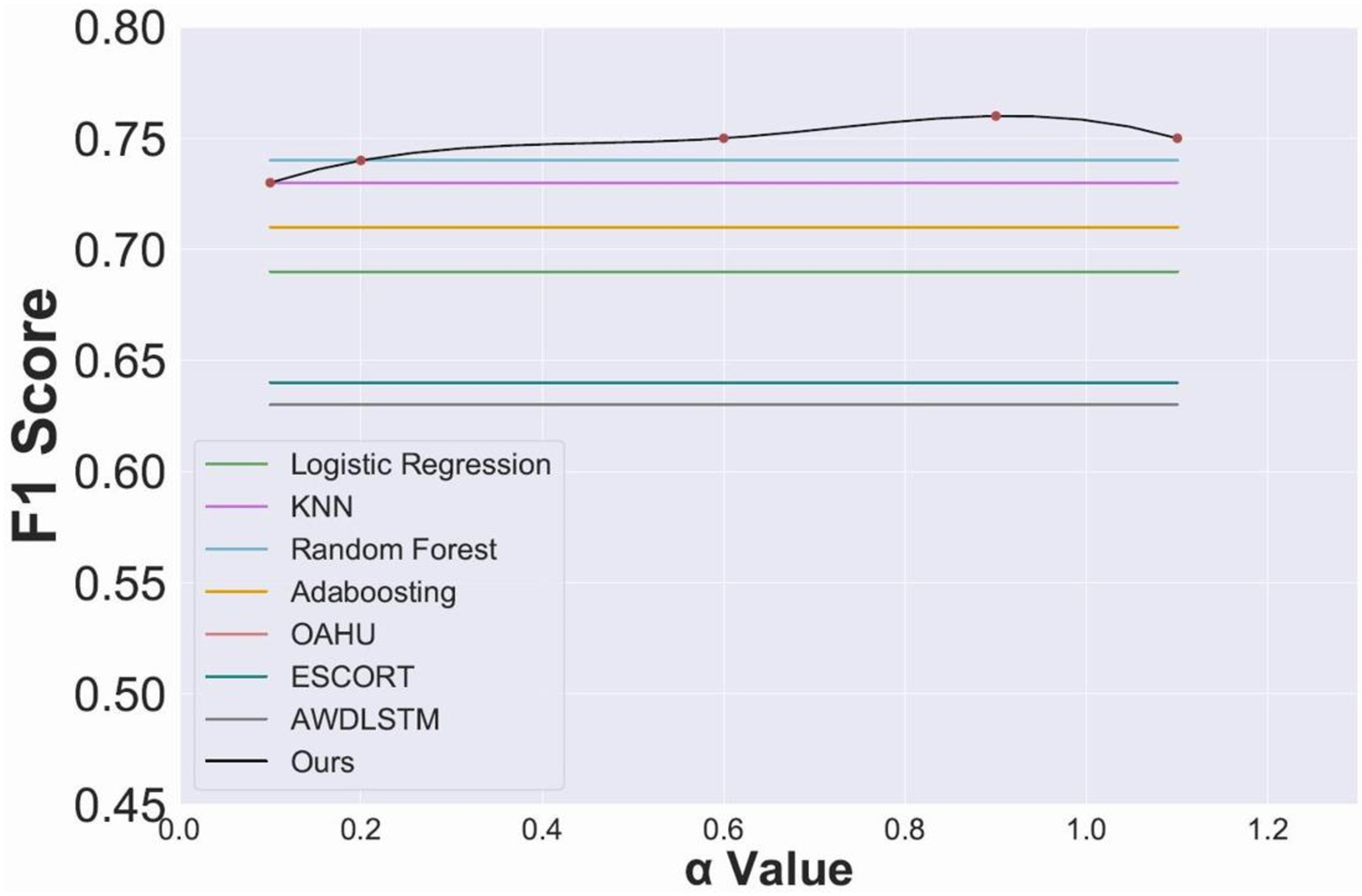}
        \end{subfigure}
        \caption{
Sensitive Analysis on the F1 Score under different epochs and $\alpha$ Value in  Eq.~\ref{eq:combine_loss}
 }
\label{fig:SensitiveAnalysis}
\end{figure*}



\begin{figure}[t]
    \centering
    \includegraphics[scale=0.22]{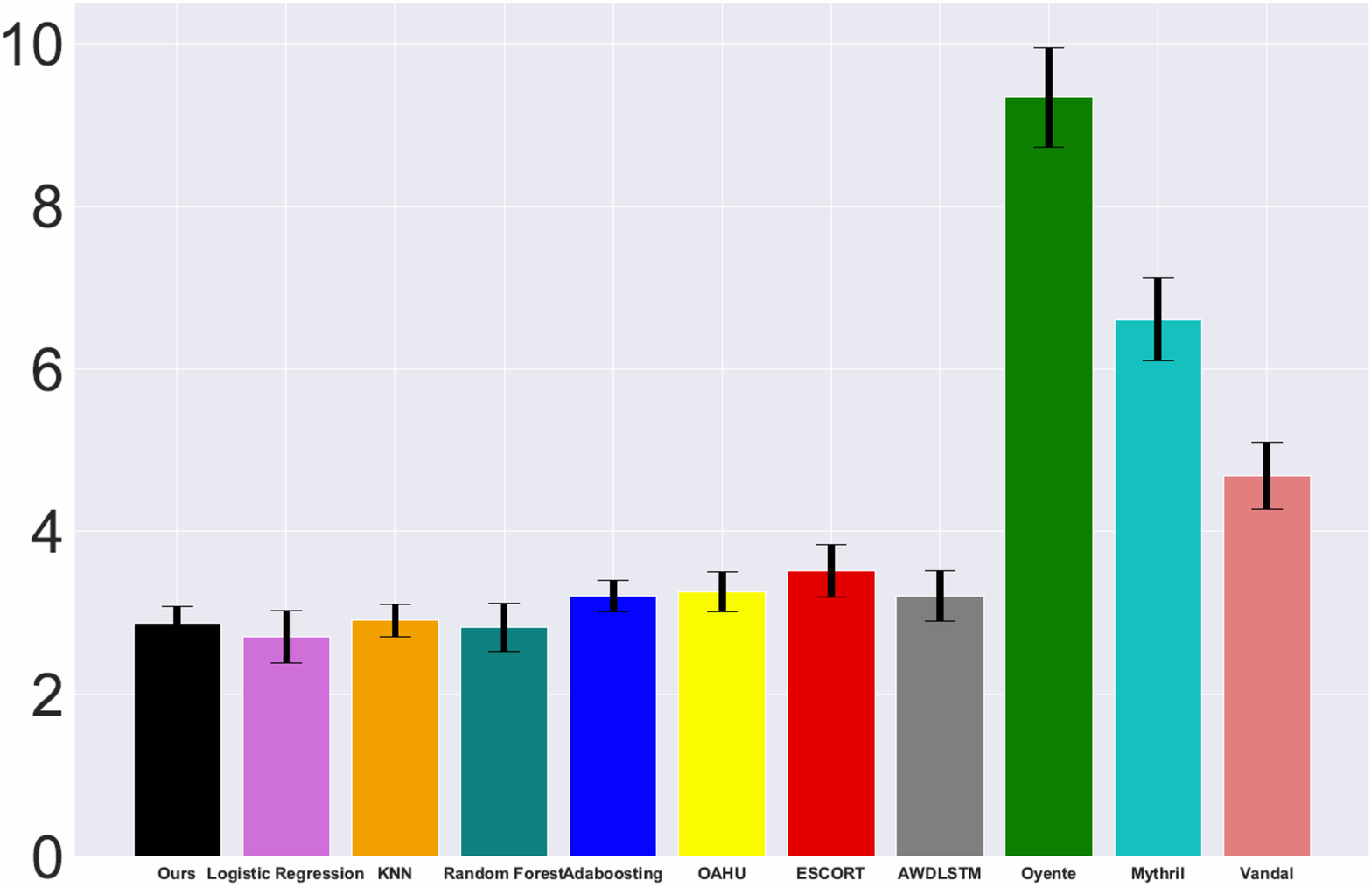}
    \caption{Detection Time Evaluation(seconds)  }
    \label{fig:fnr}
\end{figure}

\subsubsection{Smart contracts encoding methods comparisons}

To compare the performance of alternative encoding methods in \ref{sec:AlternativesFeature Matrix}, we get the feature matrix formed by these methods individually and then feed them into our classification model to compare the performance.
In Table \ref{Tab:encoding}, our method (Unigram and Bigram + TFIDF) achieves the best results with 89\% accuracy, 87\% precision, 68\% recall, 76\% F1 score, and 88\% weighted-F1 score. Doc2vec has second high results compared to other potential encodings. Our method (Unigram and Bigram + TFIDF) reflects the inter-relationship between opcodes. It applies to a new semantic (\textit{i.e.,} opcodes of smart contracts) with no additional effort. The performance of the other three methods is not as good as ours. The reasons could be: (1) Integer encoding is less useful for showing the inter-relationship between opcodes due to the integers are randomly assigned; (2) The pre-trained Longformer model may not be suitable for the smart contract domain; (3).Unigram+TFIDF form fewer features which are not enough for the classification model to learn and make detections.

\subsubsection{Detection Time Evaluation}
\label{sec:DetectionTime}
We randomly choose 2000 smart contracts in our dataset to assess the time for detection. Figure \ref{fig:fnr} shows the average time for each tool to make a decision per smart contract. Our framework spends 2.87s on average, which is significantly faster than non-machine learning-based tools(Oyente, Mythril, and Vandal). It demonstrates the efficiency of the framework. Processes such as performing disassembling, constructing CFG, and doing DFS take most of the time, which is 2.86s. Constructing feature matrix and deep learning model predicting merely cost 0.011s and 0.00019s. Other machine learning-based methods also take a similar prediction time. It is worth noticing that we did not use GPU for evaluating detection time.
\subsubsection{Unknown vulnerability detection.} \label{sssec:unkno}For rule based frameworks, when predefined rules do not cover certain types of vulnerability, they may never make detection. While our framework shows the potential to detect unknown vulnerabilities, here, we subset previous data so that training and test sets have different types of vulnerabilities. Specifically, in the training set, we randomly choose 7,200 benign smart contracts and 2,400 smart contracts suffering from multiple types of vulnerabilities but not reentrancy vulnerability. In the testing set, we randomly choose 2,400 benign smart contracts and 800 smart contracts suffering from reentrancy vulnerability. 

In Table \ref{Tab:unknown}, all the machine learning models achieve a relatively good accuracy which is around 80\% considering that the reentrancy vulnerability in the testing set has not been trained before. A fairly number of smart contracts from malicious class are correctly classified. We can conclude that the patterns from the benign class and other vulnerabilities help the model find unknown vulnerability behaviour. Although they are given smart contracts containing unknown vulnerabilities, they can still detect because the patterns differ from benign class. Furthermore, our model outperforms the other methods concretely. It achieves 86\% accuracy, 97\% precision, 44\% recall, 61\% F1 score and 84\% weighted-F1 score. We can see that our method has a deeper understanding for the benign class and known vulnerabilities than other methods and have a better performance in detecting unknown vulnerabilities. This happens due to the usage of metric learning along with DNN. Recall that metric learning strives to put same class instances together and pushes dissimilar class instances far away in the latent space. There is a possibility that a unknown vulnerable instance will be closer to the known vulnerable instances in the latent space and the model can easily uncover this unknown vulnerability with higher confidence.

\begingroup
\setlength{\tabcolsep}{1pt} 
\renewcommand{\arraystretch}{1.2} 
\begin{table}
	\centering
	\caption{Unknown Vulnerability Detection Experiment}
	\resizebox{\columnwidth}{!}{%
	\begin{tabular}{|c|c|c|c|c|c| } \hline
		\textbf{Model}& \textbf{Accuracy}& \textbf{Precision} & \textbf{Recall} & \textbf{F1 score } & \textbf{Weighted F1 score} \\ \hline
		Logistic Regression &  0.75 &  \textbf{1} &  0.01 &  0.02 & 0.65\\ \hline
		KNN (k=3) & 0.79 & 0.98 & 0.18 & 0.30 & 0.73\\ \hline
	    Random Forest  & 0.77 &  \textbf{1} & 0.07 & 0.14 & 0.68 \\ \hline
	    Adaboosting &  0.81 &  0.97  &  0.24 &  0.38 & 0.76 \\ \hline
	    OAHU \cite{gao2019towards}  &  0.74 &  0.46  &  0.03 &  0.06 & 0.65 \\ \hline
	    ESCORT \cite{escort}  &  0.79 &  0.72  &  0.43 &  0.54 & 0.77 \\ \hline
	    AWDLSTM \cite{AWDLSTM}  &  0.80 &  0.74  &  0.43 &  0.55 & 0.78 \\ \hline
		\textbf{Ours}  & \textbf{0.86} & 0.97 & \textbf{0.44} &  \textbf{0.61} & \textbf{0.84} \\ \hline

	\end{tabular}
	}
	\label{Tab:unknown}
\end{table}



\subsubsection{Ablation Study and Sensitive Analysis}
\label{sec:Ablation}
We conduct ablation study on the Majority dataset to illustrate the effectiveness of combining loss functions by visualizing data representation in the embedding space. Specifically, we first make the training data feed into these three models and keep all parts of these models the same as mentioned in Section~\ref{sec:network_configuration} except for the loss function. For the loss functions, we set $\ell = \ell_{PC}$,  $\ell = \ell_{CE}$ and $\ell = \ell_{CE} + 0.8\times \ell_{PC}$ respectively. After training, we use around 30 percent of test data and project its representation in the embedding layer on 2D space using t-SNE \cite{vanDerMaaten2008tsne}. The result is given in Figure~\ref{fig:visu}. When only use the Pairwise contrastive function $\ell_{PC}$, green dots(vulnerable class) form in one cluster, but the distance between each dot is far. In contrast,  when only using Cross-Entropy $\ell_{CE}$, green dots are not in one cluster, but green dots within each cluster are close to each other. By combining the loss function, green dots are in one cluster and are closely located to each other. The projection shows that adopting the combining loss functions approach makes the data points belonging to the same class locate closely and reduces the intra-class variances, meaning good feature representation is generated. Its Precision, Recall, and F1 scores are 0.87, 0.68, and 0.76, respectively, also better than the other two.

In Figure \ref{fig:SensitiveAnalysis}, some sensitive analysis on the performance with different number of epochs, and $\alpha$ in Eq.~\ref{eq:combine_loss} which is the weight to balance $\ell_{CE}$ and $\ell_{PC}$ . It shows that our model performs well after 200 epochs of training. When trained with more than 300 epochs, the model gradually converged. $\alpha=0.8$ works well compared to other settings. For $\alpha$ larger than 1.0, the performance begins to drop slowly.


\section{Limitations and Future Work}
\label{sec:smart_discussion}
Our framework is designed for analyzing smart contracts using binary representations. Unlike other approaches for general software applications \cite{Grieco2016,surveylin}, our framework can have limitations for those programs using binary representation. Our proposed framework can tell if a smart contract contains vulnerabilities or not but cannot pinpoint where the vulnerability lies. This limitation originates from the nature of bytecode since bytecode is less meaningful than source code. In future works, we are going to extend our framework to make more precise detection, especially the detection under unseen attack pattern \cite{unknown2019}. Also, instead of focusing on smart contracts vulnerability detection only, we will explore the possibility of conducting the label learning \cite{noiselabel} to solve the noise labeling issues in the current smart contract world.

\section{Conclusion}
\label{sec:smart_conclusion}
In this study, we proposed a novel framework utilizing metric learning based DNN to make vulnerability detection for smart contracts. 
We generate a novel feature matrix to represent smart contracts by CFG extraction and evaluate different potential encoding methods. The metric learning-based DNN shows great performance compared with other existing models. We demonstrated the effectiveness and efficiency of the framework by performing large-scale empirical experiments. 


\bibliographystyle{IEEEtran}
\bibliography{references}

\end{document}